\documentstyle[aps, preprint]{revtex}

\begin{document}

\title{Stick-Slip Fluctuations in Granular Drag}
\author{I. Albert$^{1}$, P. Tegzes$^{1,2}$, R. Albert$^1$, J. G. Sample$^1$, \\ A.-L. Barab\'asi$^{1}$, T. Vicsek$^{2}$, 
and P. Schiffer$^{1,2,*}$}
\address{
$^1$ Department of Physics, University of Notre Dame, Notre Dame, IN 46556\\
$^2$ Department of Biological Physics, E\"otv\"os University, Budapest 1117, Hungary\\
$^3$ Department of Physics, Pennsylvania State University, University Park, PA 16802
}

\date{\today}
\maketitle
\begin{abstract}
We study fluctuations in the drag force experienced 
by an object moving through a granular medium.  
The successive formation and collapse of jammed states give a 
stick-slip nature to the fluctuations which are periodic at 
small depths but become "stepped" at large depths, a transition 
which we interpret as a consequence of the long-range nature 
of the force chains and the finite size of our experiment.
Another important finding is that the mean force and the fluctuations 
appear to be independent of the properties of the contact surface between 
the grains and the dragged object. These results imply that 
the drag force originates in the bulk properties of the granular sample.
\end{abstract}

\pacs{PACS numbers: 45.70.-n, 45.70.Cc, 45.70.Ht}

\tightenlines
\section{Introduction.}

The propagation of stress in granular media presents a complex problem
which has been the subject of extensive study. An applied external
stress results in the development of an internal structure resisting the
stress, a so-called "jammed state" that is dependent on the direction
and magnitude of the stress \cite{MECates,AJLiu}. The origins of this jamming
lie in the fact that the forces do not propagate uniformly through the
granular sample but are localized along directional force chains
\cite{CLiu,BMiller,XJia,EKolb,SNCopper,AVTkachenko,LVanel,Wieghardt}, and
the jammed state is ultimately characterized by the properties of the
network of these chains. 

In this paper we investigate the jamming of spherical granular media by
analyzing the drag force opposing the movement of a solid object through
such media. The object's motion is opposed by jamming of the grains in
front of the object, but if the applied force exceeds a critical
threshold corresponding to the strength of the jammed state, the object
moves ahead while displacing the surrounding grains. Such successive
breakdowns result in large fluctuations in the drag force at low
velocities which reflect the strength of successive jammed states. Our
results demonstrate that the forces arising from such jamming of the
grains and subsequent bulk grain reorganization dominate the drag
process while the frictional forces at the surface have little effect.
Furthermore the nature of the fluctuations, and thus the strength of the
jammed states, can be affected by the finite size of the containing
vessel due to the long-range nature of the force chains. Some of the
results have been published previously \cite{IAlbert}.

\section{Description of the Apparatus}

The experimental apparatus, 
shown in Fig.\ \ref{Apparatus}, 
consists of a vertical steel cylinder of diameter $d_c$ inserted to a depth $H$ in a 
cylindrical container filled with glass spheres. A comb made of three 
5 mm diameter steel rods separated by 20 mm gaps is also inserted at 150 mm depth
opposite the vertical cylinder with the role of 
randomizing the medium.

The container  rotates with constant angular speed while the vertical 
cylinder is attached to an arm that may rotate freely around the rotational 
axis of the container.
Opposing the cylinder we have a fixed  precision
force cell \cite{ForceCell} which measures the force $F(t)$ acting on the cylinder 
as a function of time. We also incorporate a spring of 
known spring constant, $k$, between the cylinder and the force cell
with $k$ varying between 5 to 100 N/cm. The purpose of the spring is 
twofold, it allows the force to slowly build up, and, with a suitable choice
of $k$ it dominates the elastic response of 
the cylinder and all other parts of the apparatus so that the 
nonlinear deformations will not significantly affect the results.
We vary the speed ($v$) from 0.04 to 1.4 mm/s, the depth of insertion 
($H$) from 20 to $190$ mm, and the cylinder diameter ($d_c$) from 8 to 24 mm, 
studying glass ($\rho$ = 2.5 g/cm$^3$) spheres of diameter ($d_g$) 0.3, 0.5, 0.7, 0.9, 1.1, 1.4, 1.6, 3.0 
and 5.0mm. The force is sampled at 150 Hz and the response times of the 
force cell and the amplifier are less than 0.2 ms. 

As we will discuss in detail later, the drag force experienced by the 
vertical cylinder, $F(t)$, is not constant, 
but has large stick-slip fluctuations consisting
of linear rises associated with a compression of the
spring and sharp drops (sudden decompressions of the spring) 
corresponding to the collapse of the jammed grains opposing the motion. 
In most of our measurements the angular 
deflection of the arm during the fluctuations was 
around $1^\circ$ ($\approx$ one bead diameter), therefore in subsequent 
discussions we have approximated the spatial deflection of the cylinder 
as linear. Note that these experiments are conducted in
dense static granular media as opposed to drag in dilute or 
fluidized media which have been studied both theoretically \cite{VBuchholtz} 
and experimentally \cite{OZik}. In our experiments, however, all grains 
are at rest relative to each other and the container for the vast majority 
of the time. 

During the stick process the cylinder is stationary relative to the granular 
medium moving with speed $v$. During this process the 
spring is compressed at a uniform rate. Thus the change in the force on the cylinder is 
described by $\Delta F= k \Delta x = k v t_{stick}$. 
The slip processes, during which the cylinder moves relative to the medium, 
are very short when compared 
to the duration of the stick processes $t_{slip}\ll t_{stick}$ for all depths. 
Although $t_{stick}$ changes with depth ($\Delta F$ has $H$ dependence) 
$t_{slip}$ stays approximately the same,  
and it is comparable to the characteristic inertial time $\tau_{in}=2\pi\sqrt{M/k}\approx 0.1s$. 
As long as $t_{stick}$ is significantly larger than $t_{slip}$ the stick slip 
process probes the properties of the material but if $k$ is chosen very large
so that the quantity $t_{stick} = \Delta F/kv$ is comparable to $\tau_{in}$ 
the dynamics will be dominated by the inertial properties of the apparatus.
Additionally, $k$ must be chosen such that it is much smaller than the
elasticity of the apparatus. These conditions impose certain limits for acceptable 
values of $v$ and $k$ in a somewhat similar way as with sliding surface 
friction \cite{BNJPersson}. As we will show later, however, within these 
limits the average force 
and fluctuations are not affected by the choice of either $v$ or $k$.

\section{Data Analysis}

During each fluctuation the force first rises to a local maximum (the stick process)
then drops sharply as the jammed state collapses (the slip process). As shown in Fig. 
\ref{CompareRegimes} we have identified three characteristic types of stick-slip 
motion. These three types, as explained  in detail below, are:

\begin{itemize}
\item {\bf Periodic:} The signal resembles an ideal sawtooth signal.
The distributions of maxima and minima have narrow widths that do  
not overlap within two standard deviations. We observe this behavior
for beads of diameter $d_g=$ 0.3, 0.5, 0.7, 0.9 and 1.1 mm and depths 
less than 100mm.

\item {\bf Random:} The signal resembles a random 
sawtooth signal. The distributions of maxima and minima 
overlap within one standard deviation. Only beads of diameter 
larger than 1.1 mm, $d_g=$ 1.4, 1.6, 3.0 and 5.0 mm exhibit 
this behavior. We attribute the appearance of random fluctuations
to the finite size effects associated with both the container 
size and the cylinder/bead ratio. Because of the complications 
introduced by these effects, in the present paper we 
focus on the other two regimes and we plan to study this 
random regime in more  detail in another experiment.

\item {\bf Stepped:} The force builds up as a sum of small
sawtooth like steps. Each subsequent maximum and minimum tends to be 
at a higher force than the previous one, until a large reorganization 
occurs that resets the system and the incremental buildup starts again.
This behavior can be observed at high depth, ($H > 140$ mm), 
and the transition 
from the periodic to the stepped regime can be well characterized. As we will 
show later, the container diameter has an influence on the critical 
depth at which the periodic regime changes into the stepped one.

\end{itemize}

In each of these regimes the quantity $\frac{1}{v}\frac{dF}{dt}$ in the stick processes 
stays constant and equal to $k$, thus serving as strong evidence that 
during the stick processes the cylinder is at rest relative to the granular 
reference frame. We note that we did not observe precursors to the slip 
events even for the lowest speeds (0.05 mm/s) and weakest springs (5 N/m), 
cases that offered the best time resolution. 
We believe that the gradual, plastic yield observed 
right before slipping of 2D sheared granular layers \cite{SNasuno} 
does not take place in our system. 
We have extensively studied both the periodic regime and the 
transition from the periodic to the stepped regime. 

\subsection{The Periodic Regime}

{\it 1. General characteristics.} This regime has been observed  
for all beads with $d_g<1.4$  mm and depths $H$ 
smaller than 80 mm. A characteristic dataset and power spectrum (the 
squared amplitudes of the Fourier components of $F(t)$) is shown 
in Fig.\ \ref{PeriodicSignal}. 

As seen on the spectrum, the low frequency interval is dominated by the
Fourier terms corresponding to the periodicity of the signal. We observe
only one or two higher order harmonics, while at higher frequencies the
spectra exhibit power law scaling with exponent of -2. Similar scaling
was also observed in sheared granular materials in a Couette geometry
\cite{BMiller}, and is intrinsic to random sawtooth signals
\cite{ALDemirel}. At high frequencies we see only electronic noise. 

In the periodic regime, the signal $F(t)$ is made up by 
continuous rises to a maximum  value ($F_{max}$) then drops to a 
minimum ($F_{min}$). As shown on Fig.\ \ref{PeriodicMaxMin}, 
the widths of the distributions for the maxima and minima are roughly the same and 
the histograms do not intersect. We may consider the separation between the 
mean value of maxima and minima as a parameter in distinguishing the periodic 
regime from the random one. This parameter has a purely statistical interpretation, 
$4\sigma$ separation meaning that $\approx$96\% of the minima were smaller 
than $\approx$96\% of the maxima. The size distribution of upward and downward 
jumps (the difference between the value of consecutive minima to maxima and 
maxima to minima) is quite similar, as shown in Fig.\ \ref{PeriodicUpDown}. 

We observed no significant correlation between consecutive values of maxima 
and minima, thus we may consider them independent random variables.
The interval-averaged values, computed as the average value of 
the maxima or minima over an interval (50 points), however, show a 
correlated overall modulation with an amplitude of about $\pm$5\% of 
the total value. Since this modulation affects identically both the maxima 
and the minima, it appears to be caused by small changes of the jamming 
properties of the media in different parts of the container. Such changes might 
be caused by inhomogeneities in the packing or surface level associated with 
inserting and removing the cylinder. 

{\it 2. Depth, cylinder diameter and grain diameter dependence.} In a
previous paper \cite{RAlbert} we have reported that the drag force on a
cylinder with diameter $d_c$ inserted to a depth $H$ in a granular bed
can be written as $\overline{F} = \eta\rho gd_cH^2$ where $\eta$
characterizes the grain properties (surface friction, packing fraction,
etc.), $\rho$ is the density of the glass beads, and $g$ is
gravitational acceleration. We verified this expected dependence of $F$
on $d_c$ and $H$ as shown Fig.\ \ref{PeriodicHDependance}. The power
spectra as function of both depth and cylinder diameter are also
presented in Fig.\ \ref{PeriodicMultiplePower}.

When we varied the grain diameter, $d_g$, our results were always in 
agreement with our general formula: $\overline{F} = \eta\rho gd_cH^2$.
In this formula $\eta$ only describes the surface properties of the
granular material, therefore, the average force is expected to be independent of the
grain size $d_g$. As we show in Fig.\ \ref{EtaGraph}. 
while there is some scattering
around the mean for the experimentally measured values of $\eta$ there is no 
trend corresponding to the change in grain diameter.
The period, the time interval between two peaks, is 
connected to the change in the force via the formula $T=\Delta F/kv$ and
also shows a linear dependence on cylinder size 
(Fig.\ \ref{PeriodicJumpvsDiameter}) and a quadratic dependence on depth up to $h=80$ mm.

{\it 3. Scaling with $k$ and $v$.}
Since our apparatus allowed us to vary $k$ and $v$, we first 
verified whether the overall behavior of the medium is influenced by the choice 
of these parameters. From a methodological standpoint, decreasing speed or decreasing 
$k$ affect the results in a similar way, by reducing the number of  stick-slip events 
within a certain time interval, thus increasing the time resolution of all such events.
We were able to vary $k$ between 5 N/cm to 800 N/cm, and 
we find that above  $k=100$ N/cm  the signal started to lose its 
shape and became more and more distorted with increasing $k$ (see Fig.\ \ref{KDistort}). 
  
Since the amplitude of the signal and thus the period is inversely 
proportional to $k$, increasing $k$ reduced the duration of the process 
during which the cylinder moved with the media ($t_{stick})$ to the 
point at which the inertial effects and the elasticity of the jammed medium 
and apparatus were interfering with the force measurements. 
This has set an upper limit for acceptable values of $k$. 
Below this limit,  however, we could not observe any influence of $k$ 
on the data. We verify this assumption by 
rescaling the spectra corresponding to the formula $\Delta F=kv\Delta t$. 
If $k$ and $v$ affects the data only via this relation then the 
scaled power spectra expressed as $kvP(\omega) = f(\omega/kv)$ should overlap.

As seen in Fig.\ \ref{ScalingPower},
the power spectra collapse 
very well with $k$ (for data taken with constant $v$). A similar scaling 
performed for the speed $v$, with $v$ 
varied from 0.05 mm/s to 0.5 mm/s is shown in Fig.\ \ref{ScalingSpeed} 
($k$ fixed). Based on these two results we believe that the choices of  
$k$ and $v$ did not affect the nature of the force fluctuations measured by this 
experiment, i.e. that the force fluctuations represent static properties of the
grains rather than the measurement process. 

{\it Surface friction and cylinder profile.}
We have varied the friction coefficient of the surface of the cylinders 
by using cylinders made from different materials. We tested sandblasted aluminum and steel,
smooth plexiglass and teflon coated steel cylinders. 
By this substitution we varied the static
friction coefficient by a factor of $\approx 2.5$ \cite{FrictionExperiment}. 
The mean drag force was almost identical for all of these cylinders
(within 5\%)  and the fluctuations also showed no qualitative difference.
(Fig.\ \ref{Friction}). 
 
Thus, it appears that the frictional forces at the surface of an object 
have a negligible contribution to the drag force. This implies that 
the fluctuations originate in the bulk of the medium.
It is also noteworthy that the mass of these cylinders varied significantly 
(the steel was around 3 times heavier than the aluminum cylinder) demonstrating 
that inertial effects are not significant. 

We have also tested the effects of the shape of the object on the drag force, and
we found similarly unexpected behavior. There was very little difference in 
the signal profile when we substituted our circular cylinder with a half cylinder 
with the plane of the cut perpendicular to the direction of motion. While
we do not want to suggest that the force is totally independent 
of the shape of the object we believe that the shape has a much smaller
impact than one would expect from an analogy with
the shape effects on the drag in fluids (Fig.\ \ref{RodComparison}). 
 
This relative independence of cylinder shape and surface properties can, however, 
be explained qualitatively as a consequence of the nature of force propagation
in granular media. As the cylinder 
advances it has to displace  material through dilation, thus it has 
to reorganize the grains into spatially 
different positions. Since the force chains have a long-range character, the 
volume of the material resisting the movement is much
larger than the volume of material engaged into frictional 
interactions at the surface of the object. The intergrain interactions are not only 
independent of the cylinder-grain interactions but they are also far more numerous.
Thus, it is easy to see that they may completely dominate the
dynamics. One can also describe this mechanism based on the energetics of the stick-slip 
process. The energy stored in the spring will be dissipated in friction at the surface of the 
object and in friction among grains far from the cylinder, as the work necessary to bring 
them into their new spatial positions. The number of the grains involved is not affected
by the friction coefficient of the object and is only slightly influenced by the 
shape of the object. 

\subsection{Stepped regime}

A striking feature of the data is that 
the fluctuations change character 
with depth. For smaller grains $d_g<1.6$ mm and $H < H_c \approx 80$ mm the fluctuations 
are quite periodic, i.e. 
$F(t)$ increases continuously to a nearly constant value of $F_{max}$ 
and then collapses with a nearly constant drop of 
$\Delta F$ (Fig.\ \ref{TransitionShape}) 
as discussed above.
As the depth increases, however, we observe a change in $F(t)$ 
to a "stepped" signal: instead of a long linear increase followed by a 
roughly equal sudden drop, $F(t)$ rises in small linear increments to increasing
values of $F_{max}$, followed by small drops (in which $\Delta F$ is significantly smaller than the rises, 15-20\%), 
until $F_{max}$ reaches a characteristic high value, at which point 
a large drop is observed. This transition from a periodic to a "stepped" regime
is best quantified  in Fig.\ \ref{ContainerSize}, 
where we plot the depth dependence of the mean of $\Delta F$.
In the periodic regime,  $\overline{\Delta F}$ increases with depth just as  
$\overline F$ does.
As the large uniform rises of the periodic regime are broken up by the small
intermediate drops, however, $\overline{\Delta F}$ shows a local minimum and 
then increases continuously again.
The transition can also be observed in the power spectra 
as shown in Fig.\ \ref{PeriodicMultiplePower}
For low depths the power spectra display a distinct peak
characteristic of periodic fluctuations, but 
these peaks are suppressed for $H>H_c$ in correlation with the changes in  
the qualitative character of $F(t)$. 

The transition can also be observed in the histograms 
(curves that can be well approximated as gaussians) depicting 
the distribution of maxima and minima (Fig.\ \ref{DepthHistogram}). The 
initially well-separated peaks corresponding to an ideal sawtooth pattern
widen while the distance between their maxima decreases up to the point 
at which they intersect within less than one $\sigma$ separation.

Another useful way of characterizing the structure of the fluctuations and
the transition to the stepped regime is by building 
return maps, plotting $F^{max}_N$ vs $F^{max}_{N+1}$ (Fig.\ \ref{ReturnMap}). 
In these plots, the continuous line shows the $F^{max}_{N}=F^{max}_{N+1}$ 
line, points above 
it mean that the $N+1$th maximum is larger than the $N$th one, 
while points below mean that the $N$th maximum is larger than the 
next one. 
 
As long as there is no relation between 
consecutive maxima and minima these maps should not show 
any characteristic shape and should form a circular pattern with 
gaussian spread.
Indeed at low depths we do observe such a pattern but with a 
somewhat elongated structure, the elongation being probably caused by the previously 
mentioned overall modulation of the signal. 
At high depths, however, the blob starts to show a structure associated 
with the stepped character of the motion, suggesting a history 
dependence in the way each jammed state will further evolve. 

The graphs show that at large depths the majority of the peaks are followed
by a peak with an even greater height, and the increase from one peak to the 
next one is within an interval that has a well defined upper limit.
Interestingly this limit forms a line shifted above and parallel with 
the line of $F^{max}_{N}=F^{max}_{N+1}$. Only at the highest values of $F_{max}$ do we observe 
values that are smaller, usually significantly smaller than their predecessor.
This means that the force always has to first build up to a certain value 
before a large reorganization can occur. 

\subsection{The transition from periodic to stepped fluctuations.}

The transition from a periodic to a "stepped" signal is rather
unexpected, since it implies qualitative changes in the 
failure and reorganization process as $H$ increases and
the existence of a critical depth, $H_c$. 
An explanation for $H_c$ could be provided by 
Janssen's law \cite{Janssen} which states that the average pressure (which 
correlates directly with the local failure process) should become 
depth independent below some critical depth in containers with
finite width. 
This should not occur in our container, however, which has a 
diameter of 25 cm, significantly larger than $H_c$ ($\simeq 10$cm).
Furthermore we see no deviation in the behavior of $\overline F(H)$ 
from $\overline F\propto H^2$, which depends on the pressure increasing linearly with
the depth (Fig.\ \ref{ContainerSize} inset).

In order to account for the observed transition, we must inspect how the
force chains originating at the surface of the cylinder nucleate the
reorganizations. The motion of the cylinder attempting to advance
relative to the grains is opposed by force chains that start at the
cylinder's surface and propagate on average in the direction of the
cylinder's motion. These force chains will terminate rather differently
depending on the depth at which they originate, as shown schematically
in Fig.\ \ref{Walls}. For small $H$, some force chains will terminate on
the top surface of the granular sample and the stress can be relieved by
a slight distortion of the surface. Force chains originating at large
depths, however, will all terminate at the container's walls. Since the
wall does not permit stress relaxation, the grains in these force chains
will be more rigidly held in place. According to this picture, $H_c$
corresponds to the smallest depth for which all force chains terminate
on the wall. When the cylinder applies stress on the medium, the force
chains originating at small H ($H<H_c$) reduce their strain through a
microscopic upward relaxation of chains ending on the free surface. By
contrast, the higher rigidity of force chains originating at $H>H_c$
impedes such microscopic relaxations. Thus a higher proportion of the
total force applied by the cylinder will be supported by those force
chains, enhancing the probability of a local slip event occurring at
high depths. Such a slip event would not necessarily reorganize the
grains at all depths (for example the grains closer to the surface may
not be near the threshold of reorganization), thus the slip event might
induce only a local reorganization and a small drop in $F(t)$. The large
drops in $F(t)$ would occur when force chains at all depths are strained
to the point where the local forces are close to the threshold for a
slip event. This scenario also explains why $\overline{F} (H)$ does not
change at $H_c$, since $\overline F$ is determined by the collective
collapse of the jammed structure of the system. 

According to this picture, the transition is expected at smaller depths
in smaller containers since the force chains would terminate on the
walls sooner (seen Fig.\ \ref{Walls}). Indeed, as we show in Fig.\
\ref{ContainerSize}, the transition does occur at a depth approximately
20 mm smaller when the measurements are performed in a container 2.5
times smaller (with diameter of 100 mm). Furthermore, we fail to observe
the periodic fluctuations in any grains with diameters 1.4 mm or larger,
which is consistent with the suggested mechanism, since larger grains
correspond to a smaller effective system size. 

In Fig.\ \ref{GrainSizeLowDepth} 
we show how for low depths 
the signal changes from nearly periodic  to non-periodic just by changing 
the grain diameter. Similarly, the high depth regimes that show a stepped
behavior become periodic for grains of smaller 
diameter (Fig.\ \ref{GrainSizeHighDepth}).
This result is consistent with our earlier observation that reducing the 
container size has resulted in a decrease of the $H_c$ limit. 
 
\section{Conclusions}

In summary, we have made a detailed study of the fluctuations in granular drag. 
We find that the fluctuations are strongly affected by the long-range nature 
of the force propagation in granular media. The jamming responsible for 
these fluctuations originates from a localized applied stress and 
our results point towards the need of better understanding of
how force chains originating from a point source propagate and disperse
geometrically.

We gratefully acknowledge the support of the Petroleum Research
Foundation administered by the ACS, the Alfred P. Sloan Foundation, 
NSF grants PHYS95-31383 and DMR97-01998 and NASA grant NAG3-2384.

\begin{figure}[hbt]
\caption{Schematic view of the apparatus used for measuring the drag force. The
grains were contained within a 25 cm diameter rotating bucket as described in
detail previously \protect\cite{IAlbert} a) top view b) cross-sectional view.}
\label{Apparatus}
\end{figure}

\begin{figure}[hbt]
\caption{Graphical representation of the drag force fluctuations in the 
three observed regimes, periodic, random and stepped. 
Transitions between these regimes are determined by  
parameters like grain size, insertion depth and cylinder diameter.}
\label{CompareRegimes}
\end{figure}

\begin{figure}[hbt]
\caption{The signal shape and the power spectrum in the periodic regime  
($d_c=10$ mm , $d_g=0.9$ mm and depth $H=80$ mm) . The lower graph shows a typical power 
spectrum for this regime. The dashed line has the slope of $-2$.}
\label{PeriodicSignal}
\end{figure}

\begin{figure}[hbtp]
\caption{Distribution of maxima and minima in the periodic regime.
($d_c=16$ mm , $d_g=0.9$ mm and $H=60$ mm) . The large separation 
between the peaks of the histograms for the maxima 
and minima characterizes the periodic regime.}
\label{PeriodicMaxMin}
\end{figure}

\begin{figure}[hbt]
\caption{Fluctuation (upward rise and downward step) distribution in the periodic regimes 
for $d_c=16$ mm , $d_g=0.9$ mm and depth $H=80$ mm . The upward rises correspond 
to height differences from a minimum to the next maximum. Downward steps
correspond to changes from a maximum to the next minimum (absolute values).}
\label{PeriodicUpDown}
\end{figure}

\begin{figure}[hbt]
\caption{The mean value of the drag force as a function of diameter and depth.
The force has a quadratic dependence 
on the depth and a linear dependence on the diameter of the cylinder 
($d_g=0.9$ mm) .} 
\label{PeriodicHDependance}
\end{figure} 

\begin{figure}[hbt]
\caption{Power spectra for different depths with cylinder of diameter 
$d_c$=16 mm (upper graph) and different diameters at depth 
$H=80$ mm (lower graph), $d_g$=0.9mm. Data are vertically offset for clarity}
\label{PeriodicMultiplePower}
\end{figure} 

\begin{figure}[hbt]
\caption{The drag coefficient, $\eta=<F/d_c>/\rho g H^2$, as function of bead diameter. 
These data are for $H=80$, and for each 
datapoint we averaged over 5 cylinder diameters. The 
error bars represent the standard deviation.}
\label{EtaGraph}
\end{figure}

\begin{figure}[hbt]
\caption{Dependence of the size of upward rises (minima to next maxima) 
in the force on the cylinder diameter ($d_g$=0.9 mm, $H=$80 mm).}
\label{PeriodicJumpvsDiameter}
\end{figure}
 
\begin{figure}[hbt]
\caption{The effect of increasing $k$ on the signal. At very high spring 
stiffness the nonuniform elasticity of the apparatus will dominate the 
response and the inertial effects play an increasing role $H=170$ mm , 
$d_c=10$ mm , $d_g=0.9$ mm .}
\label{KDistort}
\end{figure}

\begin{figure}[hbt]
\caption{Scaling with $k$ in the periodic regime. The spring constant 
was varied from 25N/cm to 100N/cm  with the leftmost peak corresponding 
to the weakest spring ($H=80$ mm , $d_c=16$ mm , $d_g=0.9$ mm) .}
\label{ScalingPower}
\end{figure}
 
\begin{figure}[hbt]
\caption{Scaling with $v$ in the periodic regime. The speed was varied 
from 0.05 to 0.2 mm/s with the leftmost peak corresponding to the lowest 
speed ($H=60$ mm , $d_c=16$ mm , $d_g=0.9$ mm) .}
\label{ScalingSpeed}
\end{figure}

\begin{figure}[hbt]
\caption{Effects of the surface friction of the cylinder on the mean force 
and fluctuations ($d_c=16$ mm , $d_g=0.9$ mm). The error bars represent 
the standard deviations,}
\label{Friction}
\end{figure}
 
\begin{figure}[hbt]
\caption{Comparison of the average force for different surface friction and shape ($d_g=0.9$ mm) .}
\label{RodComparison}
\end{figure}

\begin{figure}[hbt]
\caption{
The characteristic fluctuations in the drag force at 4 different values
of $H$ for $d_c=10$ mm.  Note the transition from purely periodic 
fluctuations ($H\leq 60$) to stepped fluctuations with increasing depth ($H\geq 100$). 
}
\label{TransitionShape}
\end{figure}

\begin{figure}[hbt]
\caption{
The transition from periodic to stepped fluctuations as shown through 
the  magnitude of the average drop $\overline{\Delta F}$, for two 
different container sizes: circles - large container ($D =$25mm),
triangles - small container ($D=10$ mm).
The transition occurs at smaller $H_c$ in the smaller container
($d_c= 10$ mm). The inset shows the depth dependence 
of the average drag force
for $d_g = 1.1$  mm and $d_c=10$ mm, and there is no change in slope at $H_c$.
The solid line has slope 2. 
}
\label{ContainerSize}
\end{figure}

\begin{figure}
\caption{Histograms for low depth and high depth regimes. The periodic regime
shows well separated peaks while the distributions in the stepped regime overlap within 
one standard deviation.}
\label{DepthHistogram}
\end{figure}

\begin{figure}
\caption{The $F^{max}_{N+1}$ vs $F^{max}_{N}$ return maps for 
$d_g=0.9$ mm , $d_c$=10 mm and different depths. 
The straight line is the $F^{max}_{N}=F^{max}_{N+1}$ line. 
Points above the line indicate an increasing trend of the rises,
the points below show a decrease.} 
\label{ReturnMap}
\end{figure}

\begin{figure}[hbt]
\caption{Schematic diagram of the force chains perturbed by the presence of  
the walls. While the figure is greatly simplified since the force chains 
bifurcate and follow nonlinear paths there is every reason to believe 
that their effective area has a conical shape.}
\label{Walls}
\end{figure}

\begin{figure}[hbt]
\caption{Low depth ($H < H_c=100$ mm ) signal shape for 
different grain diameters. The small grain size show a 
well defined periodicity while the large grain sizes manifest a random pattern ($d_c=16$ mm) . }
\label{GrainSizeLowDepth}
\end{figure}

\begin{figure}[hbt]
\caption{High depth ($H > H_c=100$ mm ) signal shape for 
different grain diameters. The changes due 
to the increase in grain size cause the pattern to turn from nearly periodic
to the stepped stick-slip motion ($d_c=10$ mm , $d_g=0.9$ mm) . }
\label{GrainSizeHighDepth}
\end{figure}
 

\begin{references}
\bibitem{FrictionExperiment}{
We measured the friction angle $\varphi$ for the cylinders sliding on a plate covered
with glass beads glued onto its surface.($\mu=\tan(\varphi)$).
}
\bibitem{note_rodbead}{
This is why we chose to explain the periodic - stepped transition 
in the context of container size and not grain diameter.
}
\end{references}
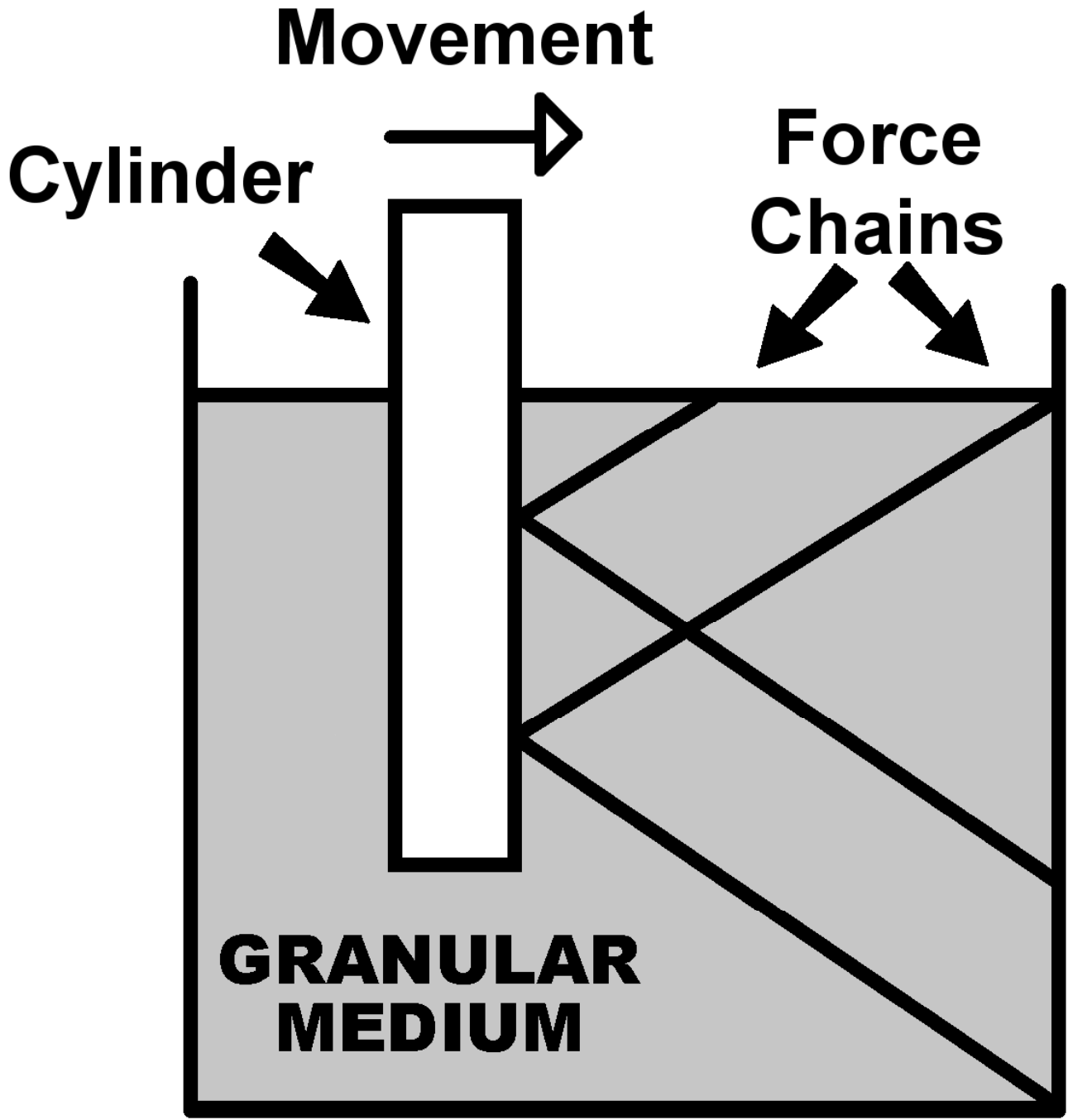
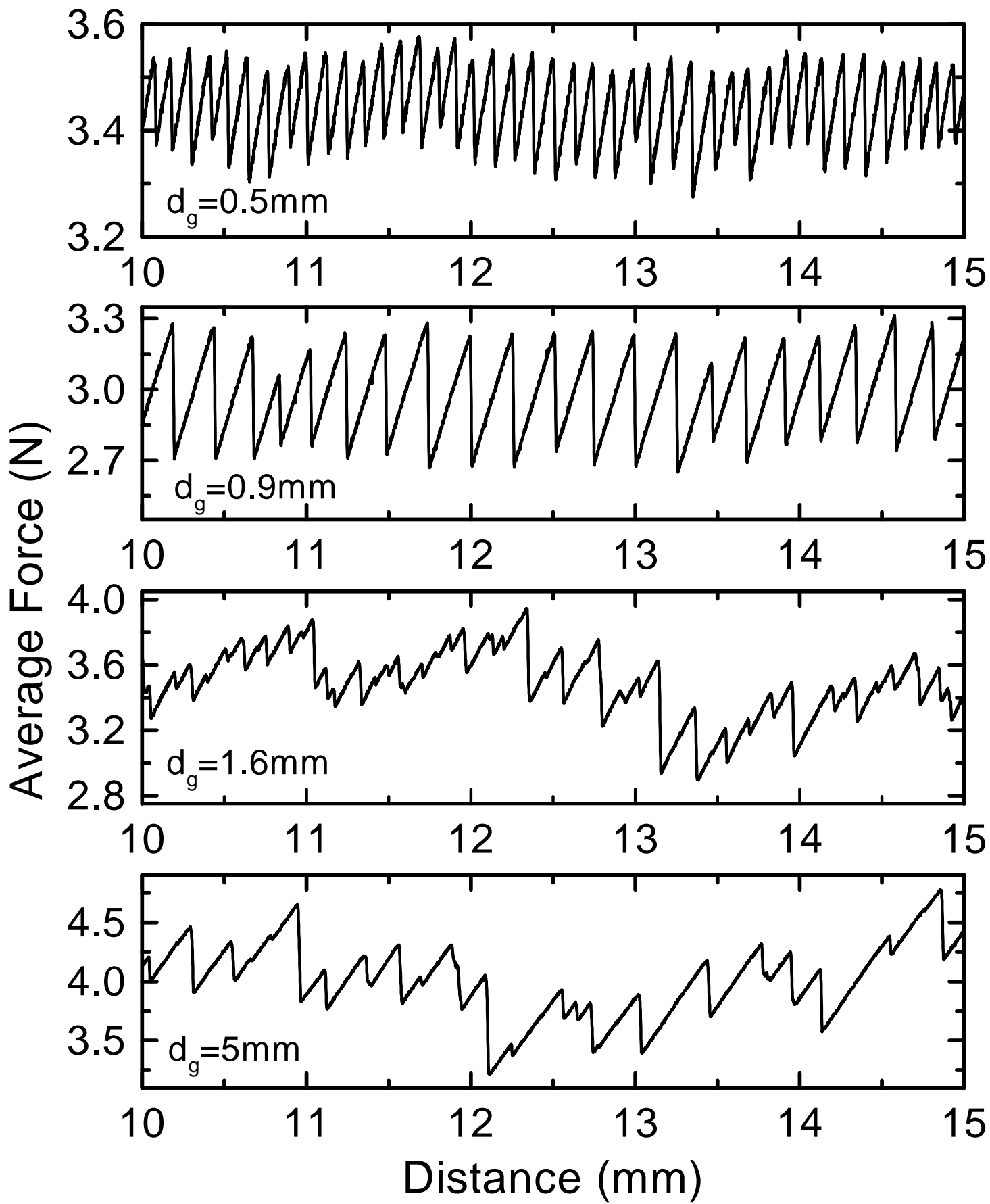
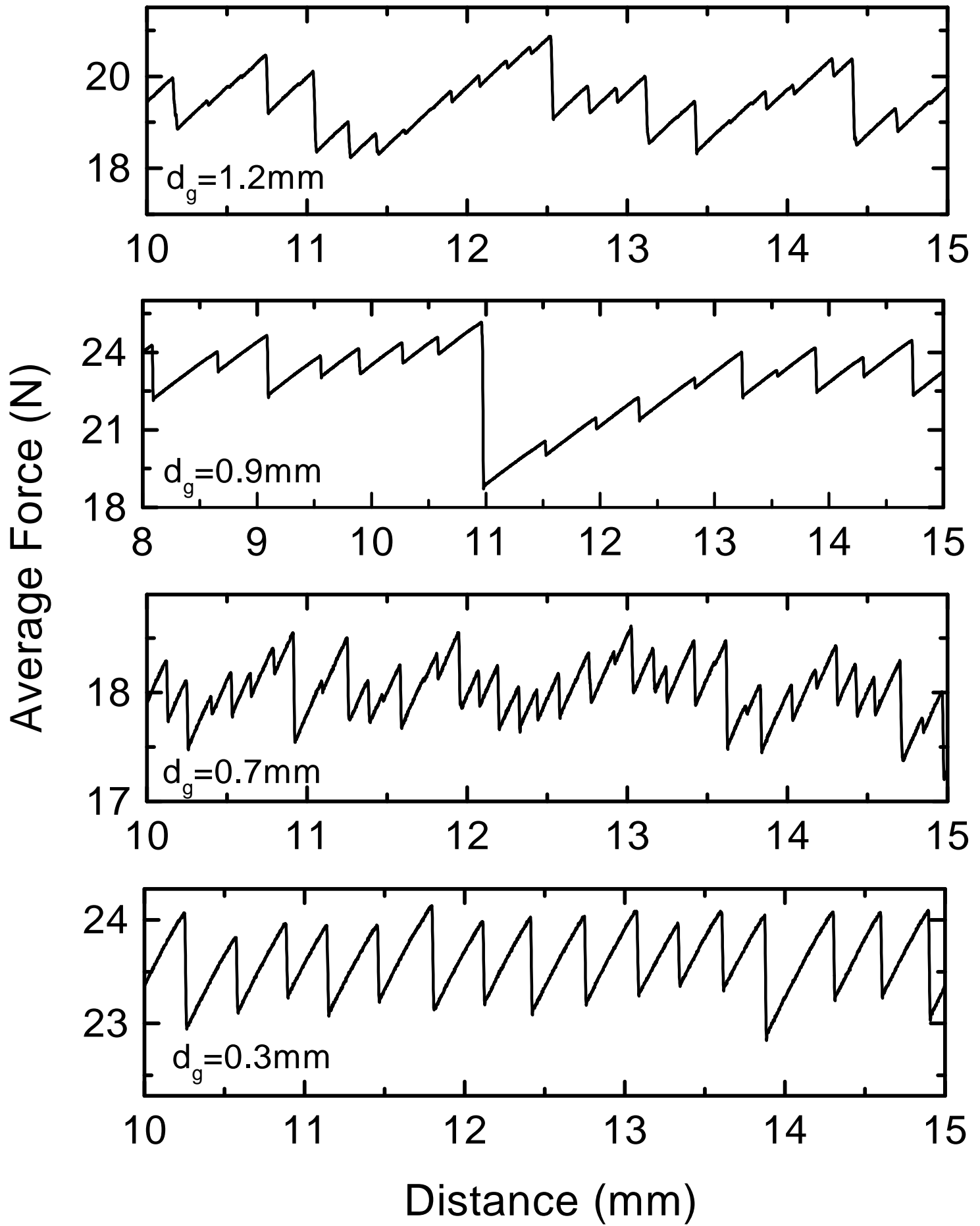
\end{document}